\documentclass[twocolumn, prl]{revtex4}

\usepackage{graphicx}% Include figure files
\usepackage{dcolumn}% Align table columns on decimal point
\usepackage{bm}% bold math

\begin{document}

\title{ Antiferromagnetism and Superconductivity in a model of Quasi 1D
      Organic Conductors} 

\author {S. Moukouri}

\affiliation{ Department of Physics and  Michigan Center for 
          Theoretical Physics \\
         University of Michigan, 2477 Randall Laboratory, Ann Arbor MI 48109}

\begin{abstract}
I apply a two-step renormalization group method to the study of the
competition between antiferromagnetism (AFM) and superconductivity in 
an anisotropic 2D Hubbard model. I show that this simple model
captures the essentials of the ground-state phases of the quasi
1D organic conductors. As found experimentally, the ground-state
phase diagram is mostly AFM. The AFM is localized in the strong-coupling
limit where the electrons are confined in the chains. It is an
SDW in the weak-coupling limit where interchain hopping is present.
There is a tiny region in the weak-coupling regime where transverse
two-particle hopping is dominant over magnetism.
\end{abstract}

\maketitle

\begin{figure}
\includegraphics[width=3. in, height=2. in]{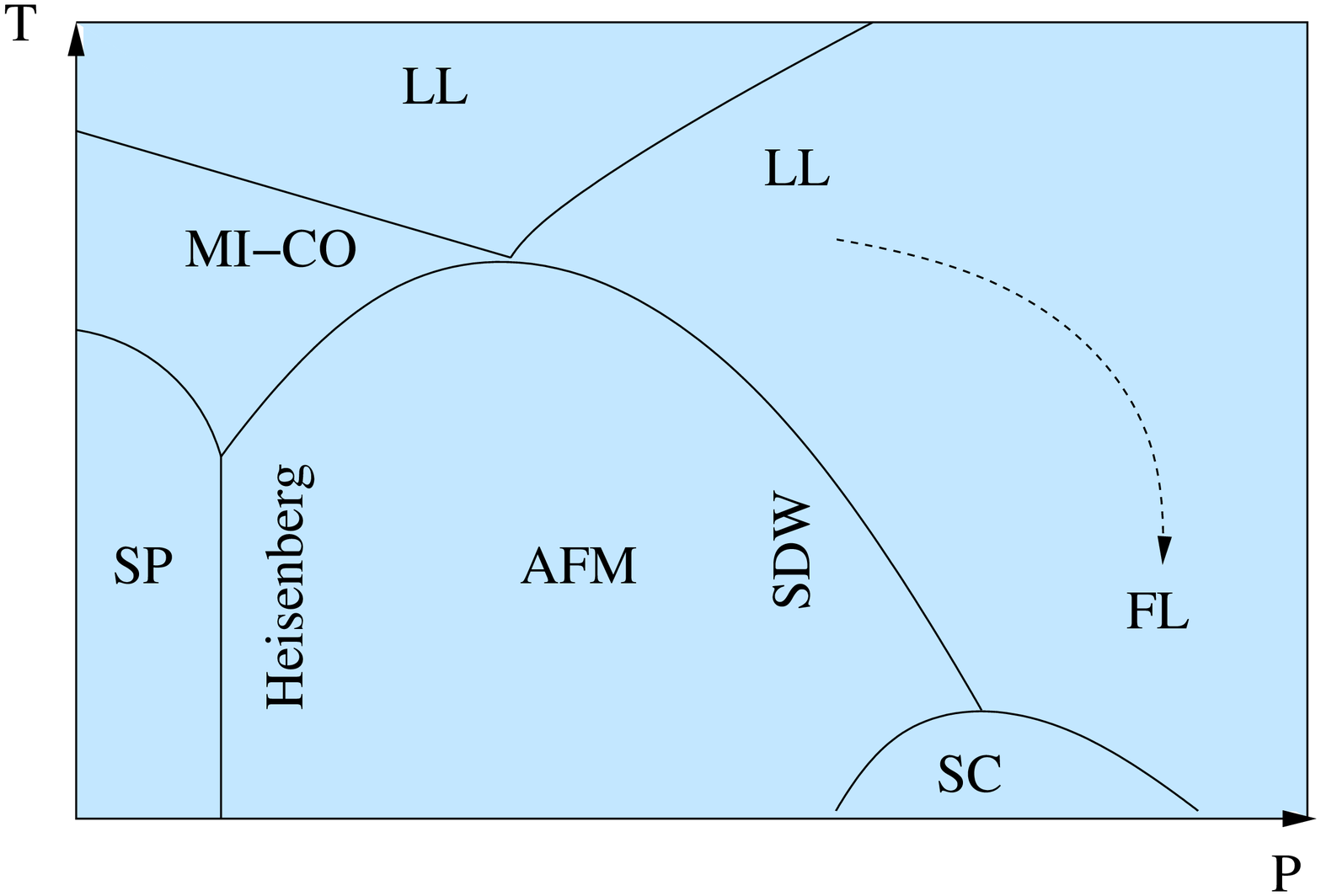}
\caption{Sketch of generic phase diagram of the quasi 1D organic
conductors: LL (Luttinger Liquid), FL (Fermi Liquid), MI (Mott Insulator),
CO (Charge Ordering), SP (Spin-Peierls), AFM (Antiferromagnet),
SC (Superconductor).}   
\vspace{0.5cm}
\label{pd}
\end{figure}

{\it Introduction.} The intriguing discovery of superconductivity (SC) lying 
next to antiferromagnetism (AFM) in the phase diagram of the charge transfer 
Bechgaard salts (see Fig.\ref{pd}) remains one of the greatest issues of 
condensed matter physics \cite{jerome-schulz}. The proximity of AFM and SC 
turned out to be a generic feature not only of the Bechgaard salt series 
$(TMTSF)_2X$, but also of the Fabre salts $(TMTTF)_2X$ and layered 2D organic 
and cuprate superconductors. In the quasi 1D organic materials, AFM occupies a 
large region of the phase diagram and is believed to be central to the 
emergence of SC. It is believed that the understanding of AFM is a 
prerequisite to that of SC.

The nature of pairing in these compounds is still unclear. Recent
experiments have yielded conflicting results. A NMR Knight shift
experiment by Lee et al. \cite{lee} found that the symmetry of the 
Cooper pairs is triplet in $(TMTSF)_2(PF)_6$. No shift was found in the 
magnetic susceptibility at the transition for measurement made under a 
magnetic field of about 1.4 {\it Tesla}. A subsequent Knight shift 
experimenent performed at lower fields reveals a decrease in the spin 
susceptibility. This result is consistent with singlet pairing.\cite{shinagawa}
The authors of this later experiment suggested a possible singlet-triplet
pairing crossover as function of the magnetic field as a resolution of
these conflicting results.

In the face of these experimental uncertainties, a theoretical input onto the 
behavior of simple models of these compounds is of crucial importance. 
A theoretical analysis of AFM in the quasi 1D organic materials was proposed 
by Bourbonnais, Caron, and coworkers  \cite{bourbonnais}. This description 
 was essentially based on a perturbative renormalization group (RG) 
applied on the g-ology model. They found that the AFM phase has two regions.
On the left side of the AFM phase, the 1D chains are Mott insulators, the
charge gap $\Delta_{\rho}$ induced by coulomb interactions is such that
$\Delta_{\rho} \gg t_{\perp}$, where $t_{\perp}$ is the transverse hopping
parameter. Hence, the single particle transverse hopping is irrelevant. 
In this region, the electrons are necessarily confined in the chains. 
$t_{\perp}$ can nevertheless generate an interchain exchange $J_{\perp}$ by 
virtual interchain hopping. Using the RG method \cite{bourbonnais-caron}, 
it can be shown that this process leads to a transverse effective Hamiltonian
$H_{\perp}=\int dx \sum_i J_{\perp} S(x)_iS(x)_{i+1}$, 
with $J_{\perp} \approx t_{\perp}^2/\Delta_{\rho}$.
In this region, the electrons are necessarily confined in the chains due to the 
irrelevance of $t_{\perp}$. As pressure increases, the electrons
progressively delocalize in the transverse direction. In the right region
of the AFM phase, the magnetism is itinerant and arises from the
nesting of the Fermi surface $\epsilon_{r,l}(k)=-\epsilon_{l,r}(k+Q)$, where
the indices $(r,l)$ stand for the right and the left parts of the Fermi
surface respectively, and $Q=(2k_F,\pi)$ is the nesting vector. The
nesting leads to the divergence of the susceptibility 
$\chi(Q,\omega=0) \propto \frac{1}{L^2} \sum_k 
\frac {f(\epsilon_r(k))-f(\epsilon_l(k+Q))}{\epsilon_r(k)-\epsilon_l(k+Q)}$.
However, this description  presents a difficulty. At a 
temperature above the SDW phase, the system is not a Fermi liquid, it is 
rather a Luttinger liquid. There are no quasi particles and the FL 
description does not apply. A perturbative RG applied
for coupled LLs can only yield the most divergent susceptibility, but it
cannot reach the ordered phase. Further increasing pressure destroys the 
magnetic order and leads to superconductivity as illustrated on the right 
part of the phase diagram. A recent RG analysis has suggested that  
superconductivity can emerge from the frustration induced by hopping between 
next-nearest neighbors chains \cite{nickel}. Pairing emerges through 
short-range AFM analogous to the Kohn-Luttinger effect induced by
Friedel oscillations\cite{emery}. 

 {\it Model and Methods.} The Hubbard model is also often used for the 
organic conductors. It
differs from the g-ology model in that it includes non-linear dispersion
and the scattering processes are not restricted to a narrow region
around the Fermi points. While it enjoys an exact solution  and can be
studied with the density-matrix renormalization group (DMRG) or the Monte
Carlo methods in 1D, it remains a difficult challenge, even numerically,
when $t_{\perp}$ is turned on. In this letter, I will show that the essential 
aspects of the ground-state phase diagram can be obtained from the single-band 
extended Hubbard model using the two-step DMRG 
method \cite{moukouri-TSDMRG, moukouri-TSDMRG2}, to which I supplement with 
a Wilson RG analysis of the low energy spectrum. I will consider the 
following model at quarter filling, the nominal density of organic
conductors, which has standard notations:

\begin{eqnarray}
\nonumber H=-t_{\parallel}\sum_{i,l,\sigma}(c_{i,l,\sigma}^{\dagger}
c_{i+1,l,\sigma}+h.c.)
 +U\sum_{i,l} n_{i,l,\uparrow}n_{i,l,\downarrow} \\
\nonumber + V\sum_{i,l,\sigma}n_{i,l, \sigma}
n_{i+1,l,\sigma} 
-\mu \sum_{i,l,\sigma} n_{i,l,\sigma} \\
 -t_{\perp}\sum_{i,l,\sigma}(c_{i,l,\sigma}^{\dagger}c_{i,l+1,\sigma}+h.c.)+
V_{\perp}\sum_{i,l,\sigma} n_{i,l, \sigma} n_{i,l+1,\sigma}.  
\label{hamiltonian}
\end{eqnarray}

The two-step RG method starts by using DMRG to compute the low energy 
eigenvalues
and eigenvectors of a single chain of lenght $L$. Then Hamiltonian 
$H$ is projected onto the tensor product of the eigenvectors 
of the disconnected chains. This leads to the effective one-dimensional 
Hamiltonian

\begin{eqnarray}
\nonumber \tilde{H} \approx \sum_{l} H_{0,l} 
- t_{\perp}\sum_{l,\sigma}(\tilde{c}_{l,\sigma}^{\dagger}
\tilde{c}_{l+1,\sigma}+h.c.) \\
+V_{\perp}\sum_{i,l,\sigma}{\tilde n}_{i,l, \sigma}{\tilde n}_{i,l+1,\sigma},
\end{eqnarray}
 
\noindent where $H_{0,l}$ is diagonal, its elements are the eigenvalues
of the single chain, the operators $\tilde{c}_{l,\sigma}$ are composite 
matrices containing the renormalized $c_{i,l,\sigma}$ for $i=1,L$.
Since $\tilde{H}$ is $1D$, it may be studied using DMRG again. 

Additional
insight in the behavior of $H$ may be gained by using the Wilson RG instead
of DMRG in the second step. The advantage of the Wilson RG lies in the
fact that the low energy spectrum can be obtained. However the Wilson
method, directly applied to ${\tilde H}$, is not accurate because all 
the terms in the transverse direction are of the same order. 
I use the same trick used by Wilson for the Kondo problem \cite{wilson}.
 $\tilde{H}$ is defined as the limit of $\tilde{H}_{\Lambda}$
when $\Lambda \rightarrow 1$. $\tilde{H}_{\Lambda}$ is given by 

\begin{eqnarray}
\tilde{H}_{\Lambda}= \sum_{l} \frac{H_{l,l+1}}{\Lambda^{(l-1)/2}},
\end{eqnarray}   

\noindent where 

\begin{eqnarray}
\nonumber H_{l,l+1}=H_{0,l}+H_{0,l+1}- 
t_{\perp}\sum_{\sigma}(\tilde{c}_{l,\sigma}^{\dagger}\tilde{c}_{l+1,\sigma}+
h.c.) \\
+V_{\perp}\sum_{i,l,\sigma}{\tilde n}_{i,l, \sigma}{\tilde n}_{i,l+1,\sigma}.
\end{eqnarray}

 It is to be noted that this is not the usual Wilson's momentum space 
discretization. The justification of this scheme rests on the fact that the
essential physics remains unchanged. 
When $\Lambda > 1$, the terms corresponding to $l>1$ act as a perturbation
on the term with $l=1$. For $\Lambda$ not too large, I expect 
$\tilde{H}_{\Lambda}$ to essentially have the same behavior as  $\tilde{H}$.
But if $\lambda \gg 1$, $t_{\perp}/\Lambda$ will be too small with
respect to the finite size energy separation and the chains will be
disconnected. It is to be remarked that this approach may also be useful
if it is embedded as a cluster solver in a chain-dynamical mean-field 
approach \cite{bierman}. 

{\it Results.} The pressure variation will be mimicked by varying the Coulomb
parameters $U$ and  $V$, while I keep $t_{\parallel}$,  $t_{\perp}$ and
$V_{\perp}$ constant in most simulations. In the regime of strong $U$ and $V$, 
the isolated chains are Mott insulators, there is a large charge gap 
$\Delta_{\rho}$ while the spin degrees of freedom are gapless. 
At $V \alt 2$ for any value of $U$,there is a insulator-metal transition
\cite{mila}.
The transverse correlation functions yield information on 
which type of order will dominate. Since the parallel correlations have 
a power law decay in absence of a gap, the most dominant transverse correlation
in a given channel will automatically lead to long-range order in that 
channel, even if it is not dominant in 1D. Before performing the analysis of
the transverse correlations, it is somewhat instructive to look at the
low energy spectrum provided by the Wilson method. This will allow to
make a qualitative comparison with the evolution under pressure with
the  prediction of the perturbative RG.

\begin{figure}
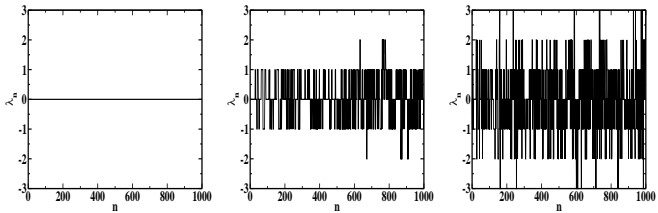

\begin{center}
$\begin{array}{c@{\hspace{0.5in}}c}
         \multicolumn{1}{l} {}\\ [-0.23cm]
\includegraphics[width=2.75cm, height=2.75cm]{renu6_6.eps}
\hspace{0.2cm}
\includegraphics[width=2.75cm, height=2.75cm]{renu4_6.eps}
\hspace{0.2cm}
\includegraphics[width=2.75cm, height=2.75cm]{renu2_6.eps}
\end{array}$
\end{center}
\caption{Charge $\lambda_n$ of lowest $1000$ excitations after 
6 RG iterations, the lattice size is $16 \times 6$, 
for $\Lambda=1.2$: in the localized AFM $U=6,~V=2$(left), the SDW 
$U=4,~V=0.85$ (center), and superconductor $U=2,~V=0$ (right)}
\label{msc}
\end{figure}

The low energy-excitation spectrum of ${\tilde H}$ was obtained from
the Wilson RG for $\Lambda \approx 1.2$ for a lattice size $16 \times 6$,
keeping $100$ states block. The lowest $1000$ excited states 
are shown in Fig.\ref{msc}. They are drastically different as pressure
is varied. In the confined regime for $U=6~V=2$ Fig.\ref{msc}, 
all the lowest states
have the charge $\lambda_n=0$. This is consistent with the fact that
since $t_{\perp} \ll \Delta_{\rho}$, the low energy behavior of
${\tilde H}$ is roughly identical to that of 
an Heisenberg model, only spin excitations are allowed. This is typically
the regime of Fabre salts at ambient pressure where a large charge gap
is observed in optical conductivity measurements \cite{schwartz}.
 By contrast, in the
SDW regime for $U=4~V=0.85$, spins excitation are still the lowest but
excitations with $\lambda=\pm 1$ now appear above them. 
Excitations with $\lambda=\pm 2$ are also observed at higher energy.
It is to be noted that because of numerical errors, excitations with 
$\lambda_n =1$ and $\lambda_n=-1$ which should normally have the same 
energy are shifted. In the regime with important superconductive correlations
$U=2~V=0$, excitations with $\lambda = \pm 2$ now appear closer to the
ground state. This is consistent with the increase of superconductive
correlations as we found below.

\begin{figure}
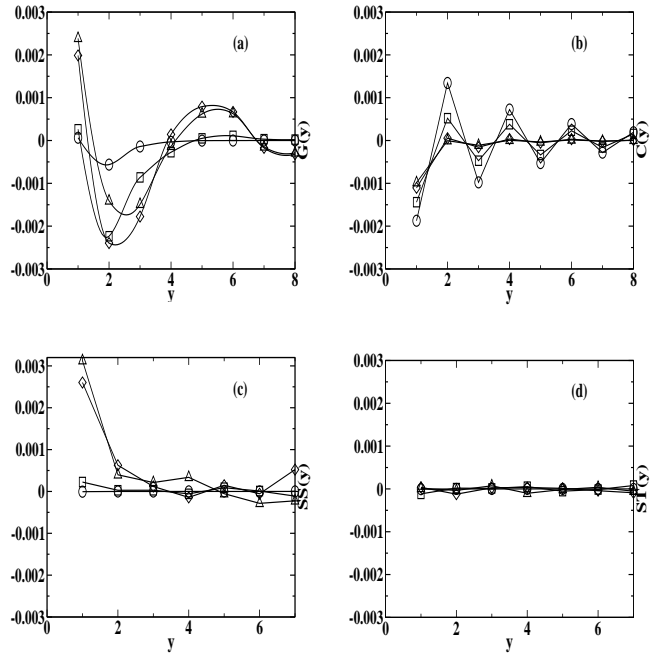

\begin{center}
$\begin{array}{c@{\hspace{0.5in}}c}
         \multicolumn{1}{l} {}\\ [-0.23cm]
\includegraphics[width=4.cm, height=4.cm]{gtp0.2.eps}
\hspace{0.5cm}
\vspace{0.35cm}
\includegraphics[width=4.cm, height=4.cm]{mtp0.2.eps}
\end{array}$
$\begin{array}{c@{\hspace{0.5in}}c}
         \multicolumn{1}{l} {}\\ [-0.23cm]
\includegraphics[width=4.cm, height=4.cm]{sstp0.2.eps}
\hspace{0.5cm}
\includegraphics[width=4.cm, height=4.cm]{sttp0.2.eps}
\end{array}$
\end{center}
\caption{Transverse interchain correlations as function of distance $y$:
 Green's function $G(y)$ (a), 
 spin-spin correlation $C(y)$ (b), singlet superconductive $SS(y)$ (c),
triplet superconductive $ST(y)$ (d) for $U=6,~V=2$ (circles), $U=4,~V=0.85$
(squares), $U=3,~V=0$ (diamonds), $U=2,~V=0$ (triangles). }
\label{gms}
\end{figure}

I now analyse the evolution of the transverse correlations when $U$ and 
$V$ are varied using the two-step DMRG for the lattice size 
$L_x \times L_y=16 \times 17$. I keep $ms_1=256$ states during 
the first step and a maximum of $ms_2=128$ states during the second step. 
For this value of $ms_2$, $\Delta E/t_{\perp} \approx 5$ which means that we
are at the limit of the two-step method as discussed in 
Ref.\onlinecite{moukouri-TSDMRG3}.
Starting from the left of the AFM phase where $U$ and $V$ are expected
to be strong, because of the presence of a large $\Delta_{\rho}$, 
the carriers are confined in the chains, even when 
$t_{\perp} \ll \Delta_{\rho}$ is turned on. The carrier confinement
was observed by Vescoli et. al in optical reflectivity measurements
\cite{vescoli}. When the oscillating electric field was oriented in the 
transverse direction, no plasma mode was observed. This carrier confinement is
seen in the behavior of the transverse Green's function 
$G(y)=\langle c_{L/2,L/2+y} c_{L/2,L/2+1}^{\dagger} \rangle$.
$G(y)$, shown in Fig.\ref{gms}(a) for $U=6$ and $V=2$, decays very
fast. $G(y) \approx 0$ for $y >3$.  This was expected given that
$t_{\perp}/\Delta_{\rho}\approx 0.1$. But as predicted by the
RG \cite{bourbonnais}, although irrelevant, $t_{\perp}$ can nevertheless
generate the motion of transverse spin degrees of freedom and lead to
magnetic order. This is seen in the transverse spin-spin correlation function 
$C(y)=\frac{1}{3}\langle {\bf S}_{L/2,L/2+y} {\bf S}_{L/2,L/2+1} \rangle$
which is shown in Fig.\ref{gms}(b). I find that despite the irrelevance
of $t_{\perp}$, $G(y)$ has its largest amplitude in the strong coupling
limit. As expected I find that the transverse singlet (triplet)
superconductive correlations 
$SS(y)=2 \langle \Delta_{L/2,L/2+y} \Delta_{L/2,L/2+1}^{\dagger} \rangle$   
($ST(y)=2 \langle \Theta_{L/2,L/2+y} \Theta_{L/2,L/2+1}^{\dagger} \rangle$),
where $\Delta_{i,l}=\frac{1}{\sqrt{2}}( c_{i,l \uparrow}c_{i+1,l \downarrow} - c_{i,l \downarrow}c_{i+1,l\uparrow})$
($\Theta_{i,l}=\frac{1}{\sqrt{2}} (c_{i,l \uparrow}c_{i+1,l \downarrow} +
c_{i,l \downarrow}c_{i+1,l\uparrow})$), are negligible in this limit
as seen in Fig.\ref{gms}(c)(d).

Moving toward the right of the phase diagram by increasing pressure
or equivalently decreasing $U$ and $V$, the carriers are expected to
progressively deconfine. In this regime, the reflectivity measurements
of Ref.\onlinecite{vescoli} reported the observation a transverse plasma
mode. The carrier deconfinement is expected for 
$t_{\perp}/\Delta_{\rho} \approx 0.5$. For $U=4$ and $V=0.85$,
the numerical simulation yields $t_{\perp}/\Delta_{\rho} \approx 0.3$
In Fig.\ref{gms}(a),(b),(c),(d), it can be seen that for $U=4$ and $V=0.85$
$G(y)$ now has a non-zero amplitude, $C(y)$ is still significant, while
$SS(y)$ and $ST(y)$ are still very small. This suggests that  the system
is in the SDW phase. The perturbative RG, coming from high temperatures,
shows that there is a 1D to 2D crossover at $T_x \approx t_{\perp}/\pi$.
At $T_x$ the 1D RG equations cease to be valid. FL arguments are used  to describe
the onset of the SDW order.  The experimental observations are however
that  both the non-FL and FL characters seem to be present depending
on the quantity measured \cite{bourbonnais}. A future application of the
TS-DMRG at finite temperature could shed light on this ineteresting
crossover regime.

If $U$ and $V$ are further reduced, $C(y)$ now decays faster despite the
fact that the amplitude of $G(y)$ is larger. $C(y)$ now vanishes for $y>3$
as seen in Fig.\ref{gms}(b) for $U=3$, $U=2$ and $V=0$ in both cases.
This implies the absence of long-range magnetic order in this regime.
At the same time, $SS(y)$ sharply increases suggesting the onset of
superconductivity in agreement with the phase diagram. The values of
$SS(y)$ at long distances are however within our margin of error.
The presence of $V_{\perp}$ is crucial to the enhancement of pairing
correlation. In the absence of $V_{\perp}$, if all other parameters
are unchanged, the dominant correlations are AFM.
$ST(y)$ is still negligeable thus suggesting the absence of triplet
pairing in the extended Hubbard model. Experimental results on the symmetry of
the pairs are still controversial. Knight shift experiments from different
groups have predicted singlet \cite{shinagawa} and triplet \cite{lee} pairings
as discussed in the introduction. This study shows a tendency towards
singlet-pairing only in the extended Hubbard model. Singlet pairing,
though interchain, was also predicted by the perturbative RG.  However,
in the RG, the pairs could be formed by carriers lying on neighboring chains
and AFM was suppressed by hopping to next-nearest neighbor chain.
It was suggested in the RG study that the interchain pairing could be 
triplet in
presence of strong enough next-nearest neighbor hopping and nearest 
neighbor Coulomb interaction both in the transverse direction. The present
two-step results do not however settle this issue. Many small effects
including longer range hopping and Coulomb interactions  
were not  included in this simple model. 
Triplet superconductivity could emerge from these terms.  

{\it Conclusion.} To summarize, I have shown that the single band extended 
Hubbard model displays the ground-state phases of the quasi 1D organic
conductors: 
(i) localized magnetism in the strong-coupling regime,
(ii) delocalized SDW magnetism in the intermediate coupling regime, and
(iii) possible superconductivity of singlet type in the weak coupling
regime. I have not discussed the spin-Peierls phase which rests at the 
extreme left of the phase diagram. In this regime,
the electron-phonon coupling is dominant over the effective transverse
exchange $J_{\perp}$, hence a pure 1D study of a spin model coupled
to phonons such as that of Ref.\cite{caron}, which shows a spin gap opening, 
captures this part of the phase diagram.

\begin{acknowledgments}
 I am very grateful to C. Bourbonnais for very helpful exchanges. 
 This work was supported by the NSF Grant No. DMR-0426775.

\end{acknowledgments}

\end{document}